\documentclass[fp,twocolumn]{jpsj3}
\usepackage{mathrsfs}
\usepackage{enumerate}
\usepackage{graphicx,exscale,amsmath,amssymb}
\def\v#1{\mib #1}
\def\dfrac#1#2{{\displaystyle\frac{#1}{#2}}}
\def\nrung{q}
\newcommand{\aver}[1]{\left\langle {#1} \right\rangle}
\def\Hlad{H^{\rm ladder}}
\def\Htub{H^{\rm tube}}
\def\ln{{\rm{ln}}}

\def\sgn{{\rm sgn}}
\def\Tr{{\rm Tr}}

\title
{
Statistical Transfer Matrix Study of the $\pm J$ Multileg Ising Ladders and Tubes}

\author
{
Kazuo {\sc Hida}
\thanks{E-mail address: hida@mail.saitama-u.ac.jp}
}

\inst
{
Division of Material Science, Graduate School of Science and Engineering, \\ Saitama University, Saitama, Saitama, 338-8570
}

\recdate
{March 2, 2012
}

\abst
{Finite temperature properties of symmetric $\pm J$ multileg Ising  ladders and tubes are investigated using the statistical transfer matrix method. The temperature dependences of the specific heat and entropy are calculated. In the case of tubes, it is found that the ground-state entropy shows an even-odd oscillation with respect to the number of legs. The same type of oscillation is also found in the ground-state energy. On the contrary, these oscillations do not take place in ladders. From the temperature dependence of the specific heat, it is found that the lowest excitation energy is $4J$ for even-leg ladders while it is $2J$ otherwise. The physical origin of these behaviors is discussed based on the structure of excitations.
}

\kword
{$\pm J$ model, multileg tube, multileg ladder, transfer matrix, ground-state energy, ground-state entropy, even-odd oscillation, droplet, domain wall
}

\begin{document}

\maketitle
\section{Introduction}
The spin systems with ladder and tube geometries have been attracting the interest from various viewpoints.\cite{ladderrev,tuberev} In the case of the spin-$1/2$ quantum Heisenberg ladders, it is well known that the properties of the ground states of the even-leg and odd-leg ladders are essentially different. Recently, the  spin tube materials are also synthesized\cite{tubeexp} and activated their theoretical studies. Various exotic quantum phenomena arising from the interplay of quantum fluctuation and frustration are predicted. 

Although the quantum spin ladders and tubes are extensively studied, their  classical counterparts have been less studied. Actually, the ground states of the regular unfrustrated classical Ising ladders and tubes are rather trivial. However, the ground state in the presence of the quenched randomness and frustration is  nontrivial even in the Ising models.\cite{mp,deri,kado}  Among them, the two-leg $\pm J$ Ising ladder is one of the simplest models with randomness and frustration.  Mattis and Paul\cite{mp}(MP) proposed the method to calculate the free energy of this model exactly, using the statistical transfer matrix method. Although their contribution was pioneering, their argument was limited to the case of two-leg ladder, and the numerical estimation of the free energy and ground-state energy was inaccurate. In this paper, we extend their method to the multileg ladders and tubes to calculate their free energy, entropy and specific heat at finite temperatures. At low temperatures, it is shown that the entropy of the classical $\pm J$ Ising tubes shows an even-odd oscillation with respect to the number of legs $q$, while the entropy of the classical $\pm J$ Ising ladder shows no oscillation. It is also found that the energy gap is $4J$ for even-leg ladders but it is $2J$ in tubes and odd-leg ladders. These features are qualitatively understood by considering the structure of the excitations.

The even-odd oscillation in the correlation length of the $\pm J$ Ising tubes was mentioned in ref. \citen{cm}. However, these authors were interested in the limit of two dimensional $\pm J$ model ($q \rightarrow \infty$) and  this effect was considered as a finite size effect which is harmful in taking the thermodynamic limit. Considering the recent increase of the interest in the models and materials with ladder and tube geometries, however, these peculiar properties of $\pm J$ Ising ladders and tubes should be investigated in more detail.

The present paper is organized as follows. In the  next section, we introduce the model Hamiltonian. The statistical transfer matrix approach by MP is extended to the multileg ladders and tubes in \S 3. The numerical results are presented in \S 4. The last section is devoted to summary and discussion. We also present the correct estimation of the ground-state energy and free energy of the two-leg ladder in Appendix.

\section{Models}
We consider the symmetric $\pm J$ Ising $\nrung$-leg ladder
\begin{align}
\Hlad &= -\sum_{n=1}^L\sum_{\alpha=1}^{\nrung-1}J_{\perp n\alpha} S_{n,\alpha} S_{n,\alpha+1}\nonumber\\
&- \sum_{n=1}^{L-1} \sum_{\alpha=1}^{\nrung}J_{n\alpha} S_{n,\alpha}S_{n+1,\alpha}\label{ham_ladder}
\end{align} 
and tube
\begin{align}
\Htub &= -\sum_{n=1}^L\sum_{\alpha=1}^{\nrung}J_{\perp n\alpha} S_{n,\alpha} S_{n,\alpha+1}\nonumber\\
&- \sum_{n=1}^{L-1} \sum_{\alpha=1}^{\nrung}J_{n\alpha} S_{n,\alpha}S_{n+1,\alpha},  \ (S_{n,\nrung+1}\equiv S_{n,1}) \label{ham_tube}
\end{align} 
where $S_{n,\alpha} (=\pm 1)$ is the Ising spin variable on the $n$-th rung and $\alpha$-th leg. The number of the rungs is denoted by $L$. The exchange constants $J_{\perp n\alpha}$ and $J_{n\alpha}$ are quenched random variables which take the values $\pm J (J > 0) $ 
  with equal probability. 

\section{Statistical Transfer Matrix Formulation}

Before constructing the transfer matrices, we gauge out the randomness along the legs from the Hamiltonians (\ref{ham_ladder}) and (\ref{ham_tube}) by the transformation
\begin{align}
\tilde{S}_{n\alpha}&=\left(\prod_{n'=1}^{n-1}\sgn{J_{n'\alpha}}\right)S_{n\alpha}.
\end{align}
The Hamiltonians (\ref{ham_ladder}) and (\ref{ham_tube}) are transformed into the forms,
\begin{align}
\Hlad &= -\sum_{n=1}^L\sum_{\alpha=1}^{{\nrung}-1}\tilde{J}_{\perp n\alpha} \tilde{S}_{n,\alpha} \tilde{S}_{n,\alpha+1}\nonumber\\
&- J\sum_{n=1}^L \sum_{\alpha=1}^{\nrung} \tilde{S}_{n,\alpha}\tilde{S}_{n+1,\alpha}, \label{ham_ladder_f}
\end{align}
and  
\begin{align}
\Htub &= -\sum_{n=1}^L\sum_{\alpha=1}^{{\nrung}}\tilde{J}_{\perp n\alpha} \tilde{S}_{n,\alpha} \tilde{S}_{n,\alpha+1}\nonumber\\
&- J\sum_{n=1}^L \sum_{\alpha=1}^{\nrung} \tilde{S}_{n,\alpha}\tilde{S}_{n+1,\alpha},  \ (\tilde{S}_{n,\nrung+1}\equiv \tilde{S}_{n,1}),  \label{ham_tube_f}
\end{align}
respectively, with
\begin{align}
\tilde{J}_{\perp n\alpha}&=J_{\perp n\alpha}\left(\prod_{n'=1}^{n-1}\sgn{J_{n'\alpha}}\right)\left(\prod_{n'=1}^{n-1}\sgn{J_{n',\alpha+1}}\right)=\pm J.  
\end{align}
We define the spin variables $\tilde{T}_{n}$ and $\tilde{L}_{n\alpha}$ by
\begin{align}
\tilde{T}_{n}&\equiv\tilde{S}_{n,1} \tilde{S}_{n+1,1}, \\
\tilde{L}_{n\alpha}&\equiv\tilde{S}_{n,\alpha} \tilde{S}_{n,\alpha+1}, \ \ (\alpha=1,...,{\nrung}-1).
\end{align}
It also follows that
\begin{align}
\tilde{S}_{n,1} \tilde{S}_{n,q}&= \prod_{\alpha=1}^{\nrung-1}\tilde{L}_{n\alpha}.
\end{align} 
Using these relations, the Hamiltonians (\ref{ham_ladder_f}) and (\ref{ham_tube_f}) are further transformed into the forms,
\begin{align}
\Hlad &= -J\sum_{n=1}^L\sum_{\alpha=1}^{{\nrung}-1}t_{n,\alpha}\tilde{L}_{n\alpha}\notag\\
&- J\sum_{n=1}^L \tilde{T}_{n} 
\left(1+\sum_{\alpha=1}^{{\nrung}-1} \prod_{\alpha'=1}^{\alpha}\tilde{L}_{n\alpha'}\tilde{L}_{n+1 \alpha'}\right), \label{ham_ladder_nloc}
\end{align} 
\begin{align}
\Htub &= -J\sum_{n=1}^L\sum_{\alpha=1}^{{\nrung}-1}t_{n,\alpha}\tilde{L}_{n\alpha}-J\sum_{n=1}^Lt_{n,q} \prod_{\alpha=1}^{\nrung-1}\tilde{L}_{n\alpha}\notag\\
&- J\sum_{n=1}^L \tilde{T}_{n} 
\left(1+\sum_{\alpha=1}^{{\nrung}-1} \prod_{\alpha'=1}^{\alpha}\tilde{L}_{n\alpha'}\tilde{L}_{n+1 \alpha'}\right), \label{ham_tube_nloc}
\end{align} 
respectively, where $ t_{n,\alpha}$'s$(=\pm 1)$ are quenched random variables. 

In this representation,  the trace over $\tilde{T}_{n}$ in the partition function $Z(\{t_{n,\alpha}\})$ can be readily taken  for each realization of $\{t_{n,\alpha}\}$. This  yields
\begin{align}
Z(\{t_{n,\alpha}\})&=\Tr_{\tilde{L}}\prod_{n} \hat{V}(\{t_{n,\alpha}\}),
\end{align} 
where  $\hat{V}{(\{t_{n\alpha}\})}$ are $ 2^{\nrung-1} \times 2^{\nrung-1} $ sized 
 transfer matrices between neighboring rungs parameterized by $\{t_{n,\alpha}\} $. In the following,  we  index the state of a rung $\{\tilde{L}_{\alpha}\}$ by $i=1+\sum_{\alpha=1}^{q-1}((\tilde{L}_{\alpha}+1)/2\times 2^{\alpha-1})$. Then, the elements of $\hat{V}$ are given by
\begin{align}
&V_{i,i'}^{\rm ladder}({\{t_{\alpha}\}})=\exp \left(\dfrac{J}{T}\sum_{\alpha=1}^{q-1} {t}_{\alpha} \tilde{L}_{\alpha}\right)\notag\\
&\times 2\cosh\left\{\dfrac{J}{T}\left(1+\sum_{\alpha=1}^{q-1} \prod_{\nu=1}^{\alpha}\tilde{L}_{\nu}\tilde{L}'_{\nu}\right)\right\}
\end{align} 
for ladders, and
\begin{align}
&V_{i,i'}^{\rm tube}({\{t_{\alpha}\}})=\exp \left\{\dfrac{J}{T}\left(\sum_{\alpha=1}^{q-1} {t}_{\alpha}\tilde{L}_{\alpha}+{t}_{q} \prod_{\alpha=1}^{\nrung-1}\tilde{L}_{\alpha}\right)\right\}\notag\\
&\times 2\cosh\left\{\dfrac{J}{T}\left(1+\sum_{\alpha=1}^{q-1} \prod_{\nu=1}^{\alpha}\tilde{L}_{\nu}\tilde{L}'_{\nu}\right)\right\}
\end{align} 
for tubes, where 
$T$ is the temperature. 

Denoting the statistical mechanical weight of the $i$-th state of the $n$-th rung by $x_{n,i}$, we define the weight vector on the $n$-th rung $\v{x}_n=(x_{n,1},x_{n,2},..,x_{n,2^{q-1}})$ with normalization $\displaystyle\sum_{i=1}^{2^{q-1}} x_{n,i}=1$. Then, the   weight vector on the $(n+1)$-th rung is determined by
\begin{align}
x_{n+1,j}&=\frac{1}{\chi_{n+1}} \sum_{i=1}^{2^{q-1}}x_{n,i}V_{i,j}(\{t_{\alpha}\})\label{eq:x}
\end{align} 
where $\chi_{n+1}$ is the normalization constant  for $\v{x}_{n+1}$ determined by 
\begin{align}
\chi_{n+1}(\v{x}_n,\{t_{n,\alpha}\})&=\sum_{i=1}^{2^{q-1}}x_{n,i}\sum_{j=1}^{2^{q-1}} V_{i,j}(\{t_{n,\alpha}\}).\label{eq:chi}
\end{align} 
It should be noted that $2^{q-1}$ ($2^{q}$ ) vectors $\v{x}_{n+1}$ and $2^{q-1}$ ($2^{q}$ )  scalars $\chi_{n+1}$ are generated from a single vector $\v{x}_n$ by the $2^{q-1}$ ($2^{q}$ ) possible choices of $\{t_{n,\alpha}\}$ in $\hat{V}^{\rm ladder}$ ($\hat{V}^{\rm tube}$ ).    

Following MP, the free energy per spin $F/N$ is expressed using $\chi_n$'s as
\begin{align}
\frac{F}{N}= -\frac{T}{N} \sum_{n=1}^L \aver{\ln \chi_{n}}_{\{t_{n,\alpha}\}}
\end{align}
where $\aver{...}_{\{t_{n,\alpha}\}}$ means the average over  $\{t_{n,\alpha}\}$ and $N(=Lq)$ is the number of spins. 

\section{Numerical Results}

\begin{figure}
\centerline{\includegraphics[width=6cm]{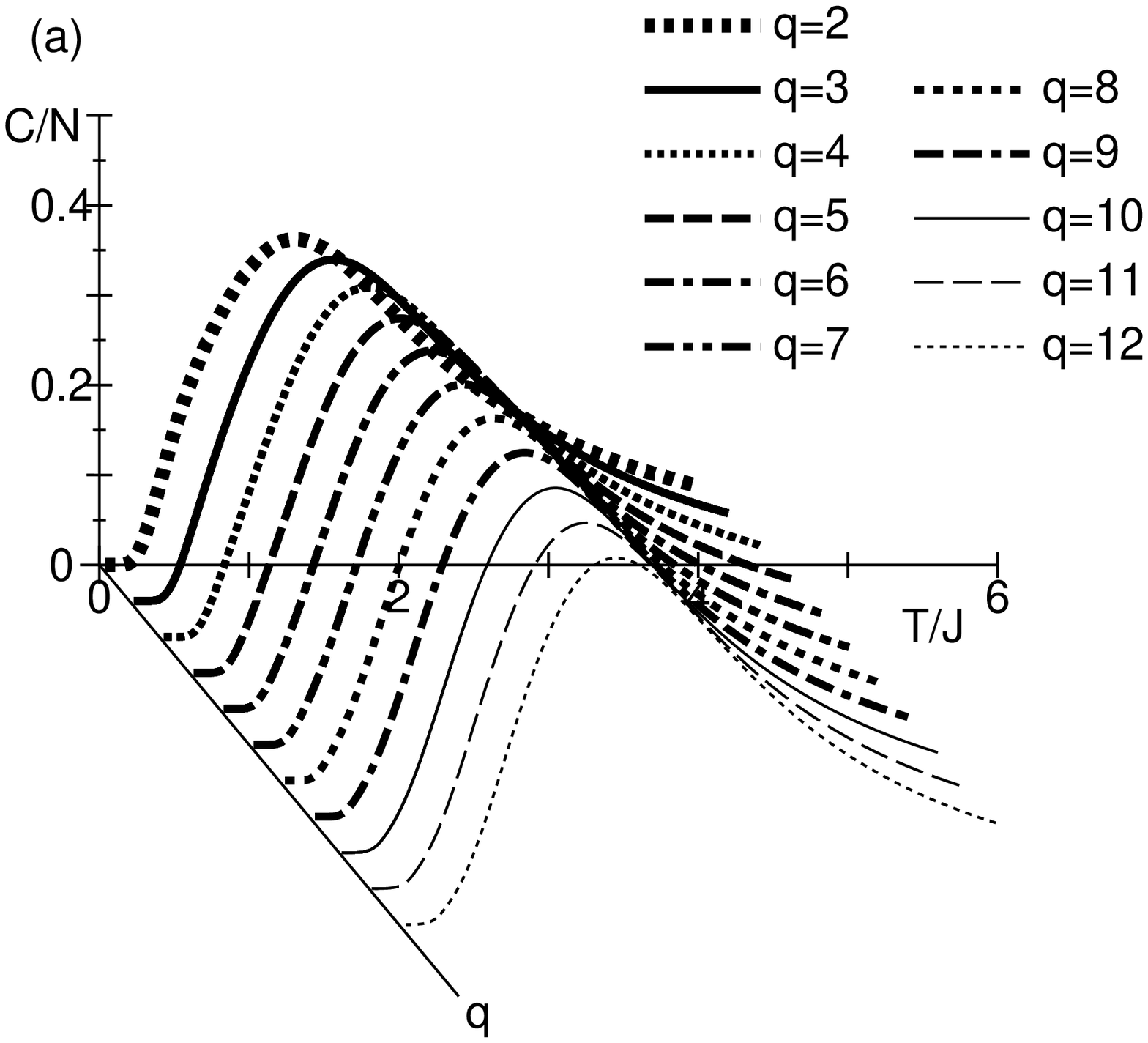}}
\centerline{\includegraphics[width=6cm]{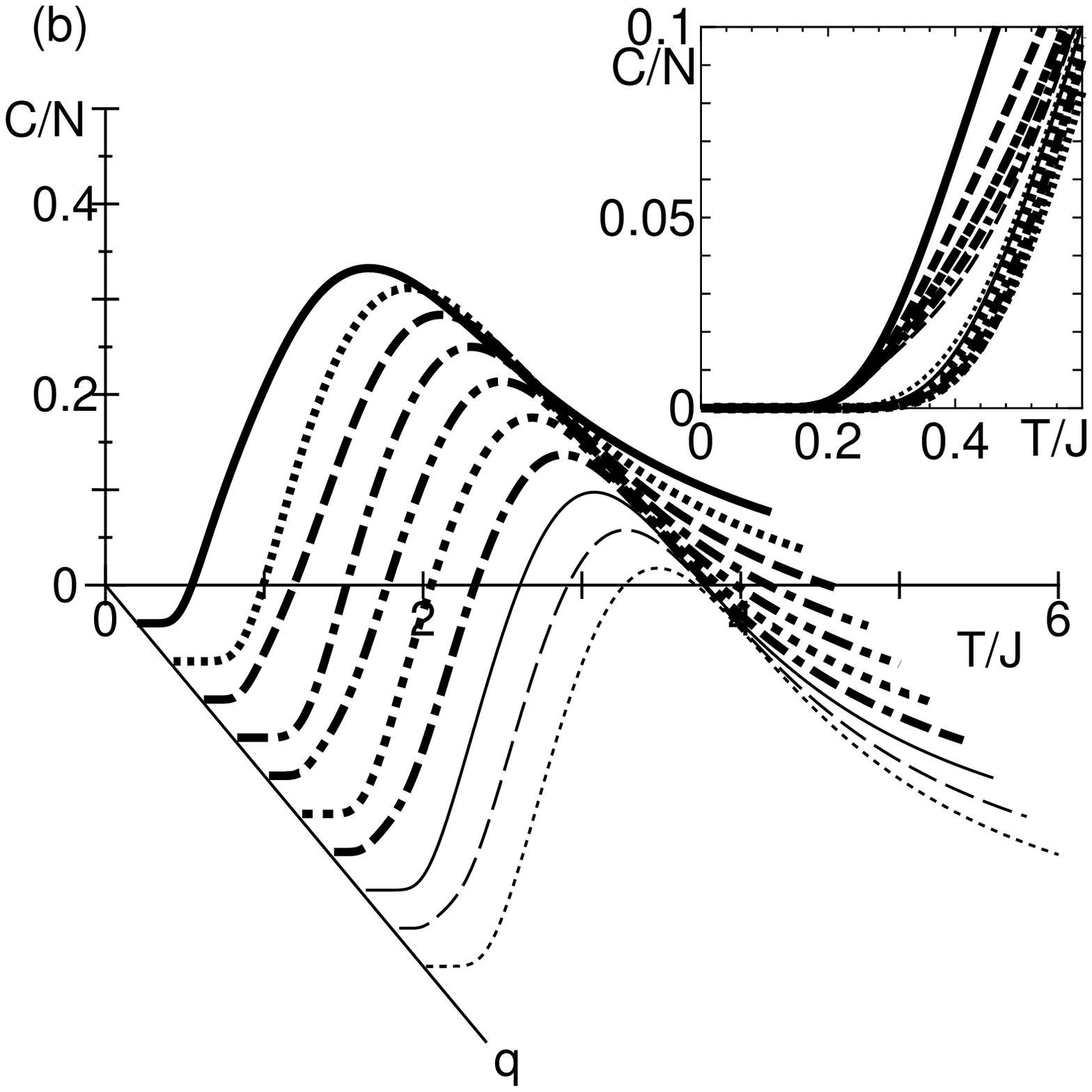}}
\caption{Temperature dependence of the specific heat per spin for (a) ladders and (b) tubes. The inset of (b) shows the temperature dependence of the latter in the low temperature regime.}
\label{hine}
\end{figure}
\begin{figure}
\centerline{\includegraphics[width=6cm]{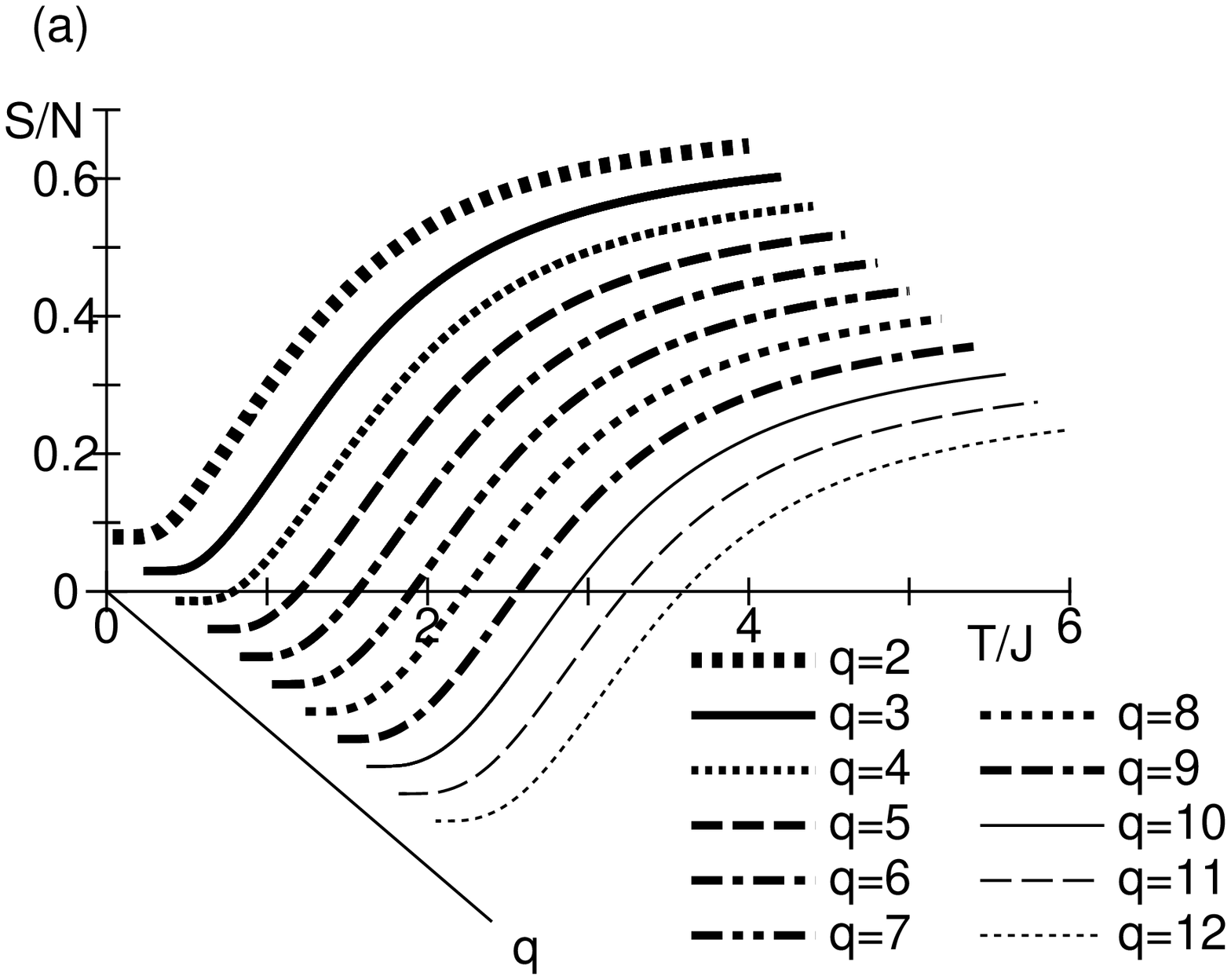}}
\centerline{\includegraphics[width=6cm]{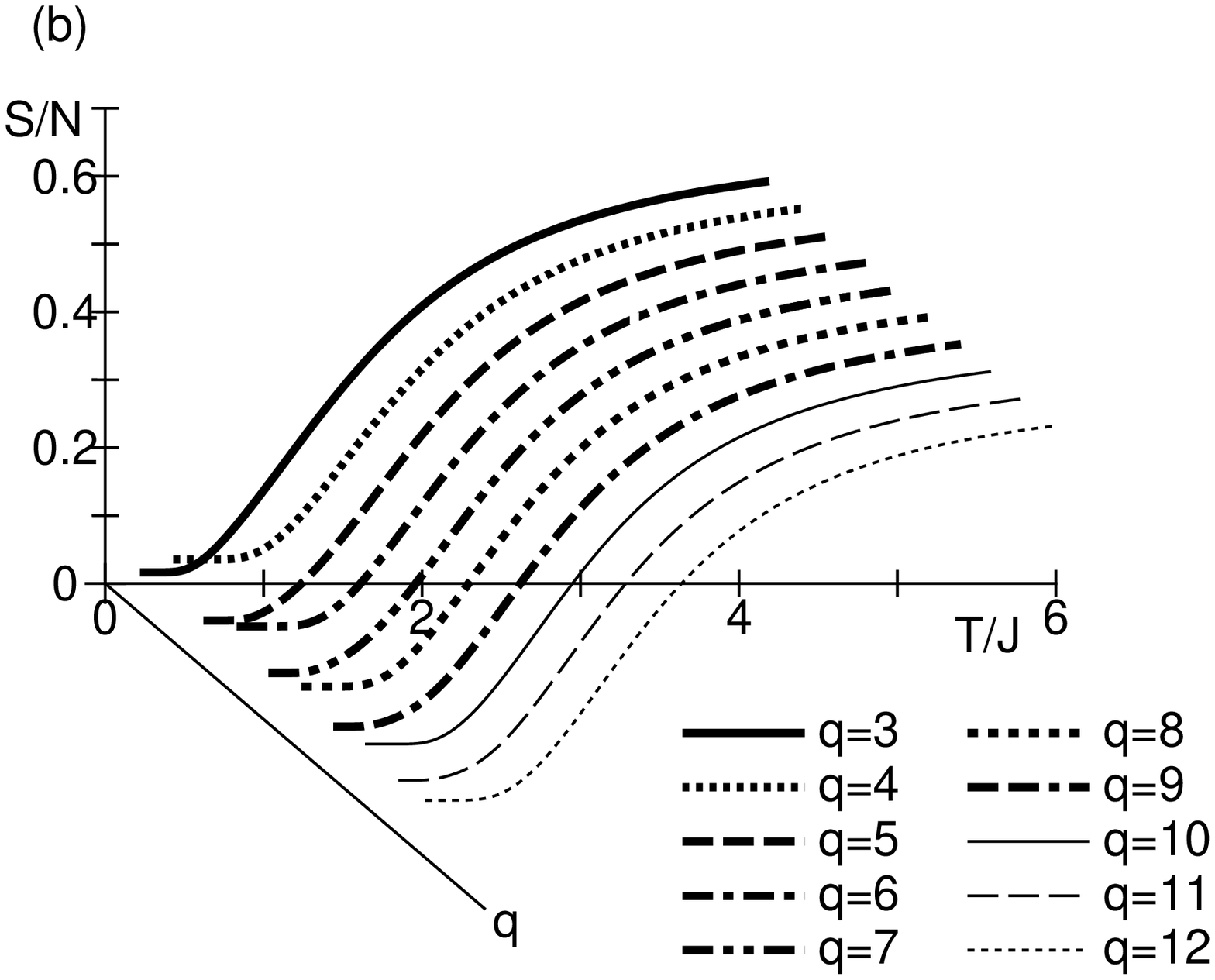}}
\caption{Temperature dependence of the entropy per spin for (a) ladders and (b) tubes.}
\label{ent}
\end{figure}
\begin{figure}
\centerline{\includegraphics[width=6cm]{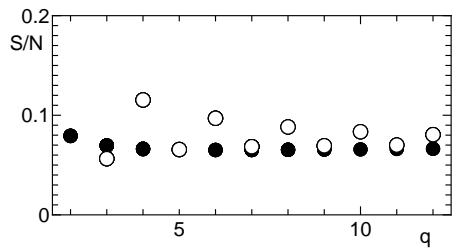}}
\caption{Ground-state entropy of ladders ($\bullet$) and tubes ($\circ$) per spin plotted against $q$.}
\label{entqdep}
\end{figure}
\begin{figure}
\centerline{\includegraphics[width=6cm]{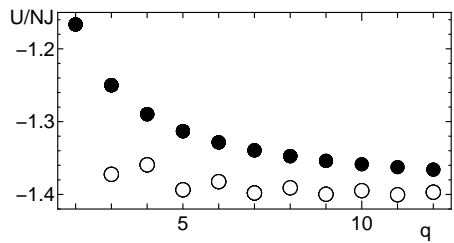}}
\caption{Ground-state energy of ladders ($\bullet$) and tubes ($\circ$)  per spin plotted against $q$.}
\label{eneqdep}
\end{figure}
\begin{figure}
\centerline{\includegraphics[width=6cm]{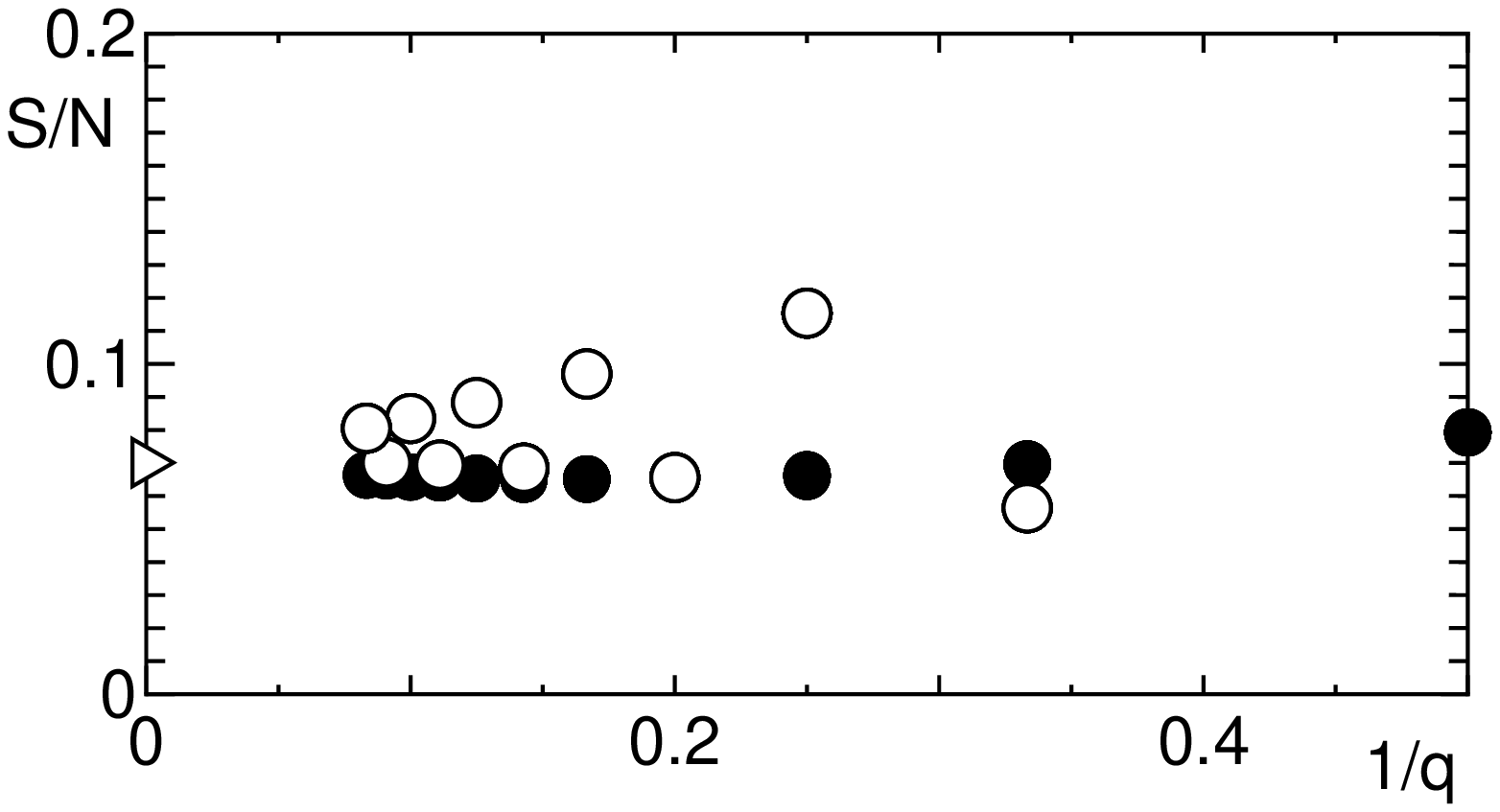}}
\caption{Ground-state entropy of ladders ($\bullet$) and tubes ($\circ$)  per spin plotted against $1/q$. The right-directed open triangle is the value for the two-dimensional $\pm J$ Ising model.}
\label{entqdepinv}
\end{figure}
\begin{figure}
\centerline{\includegraphics[width=6cm]{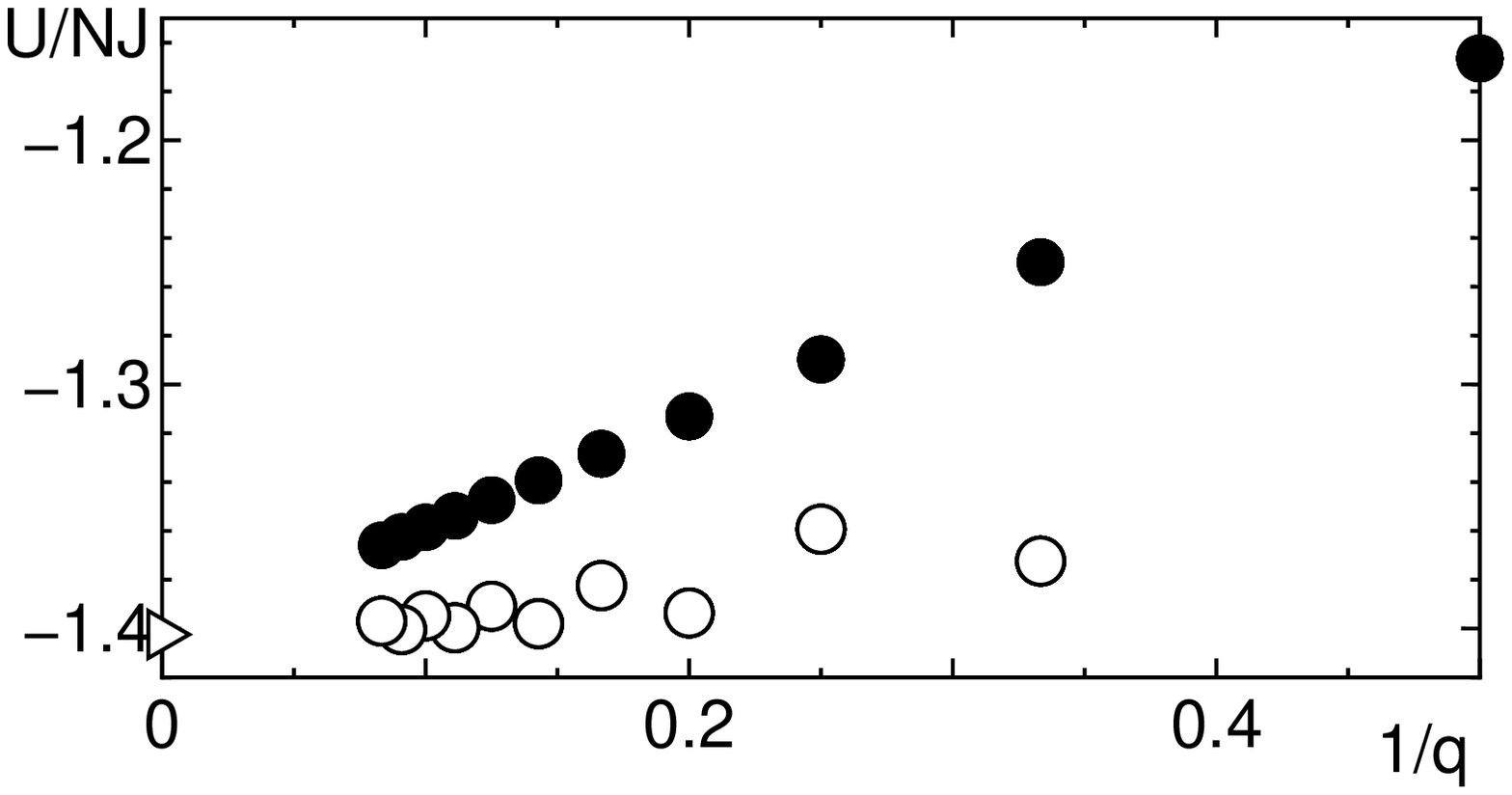}}
\caption{Ground-state energy of ladders ($\bullet$) and tubes ($\circ$)  per spin plotted against $1/q$. The right-directed open triangle is the value for the two-dimensional $\pm J$ Ising model.}
\label{eneqdepinv}
\end{figure}
\begin{figure}
\centerline{\includegraphics[width=6cm]{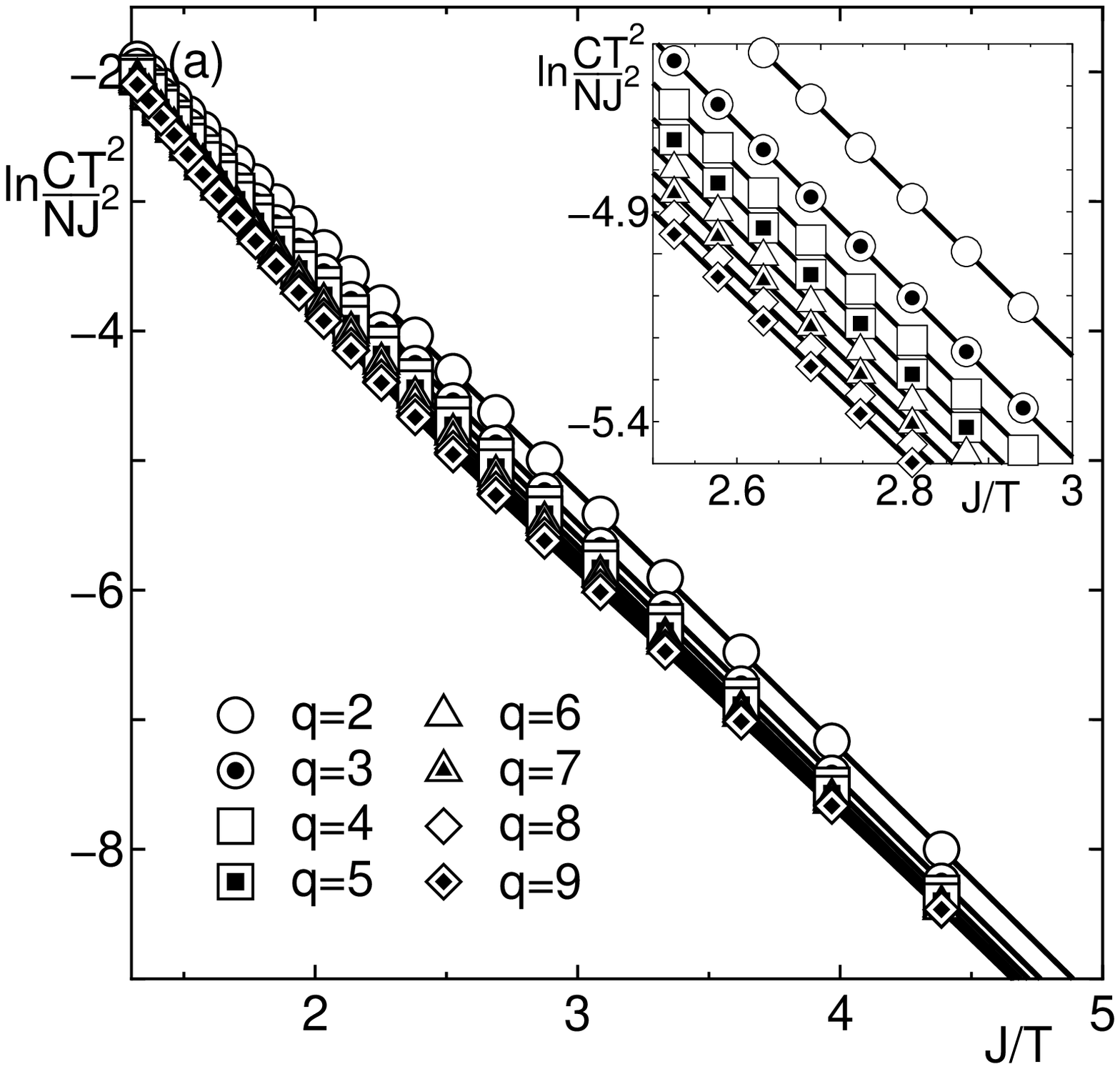}}
\centerline{\includegraphics[width=6cm]{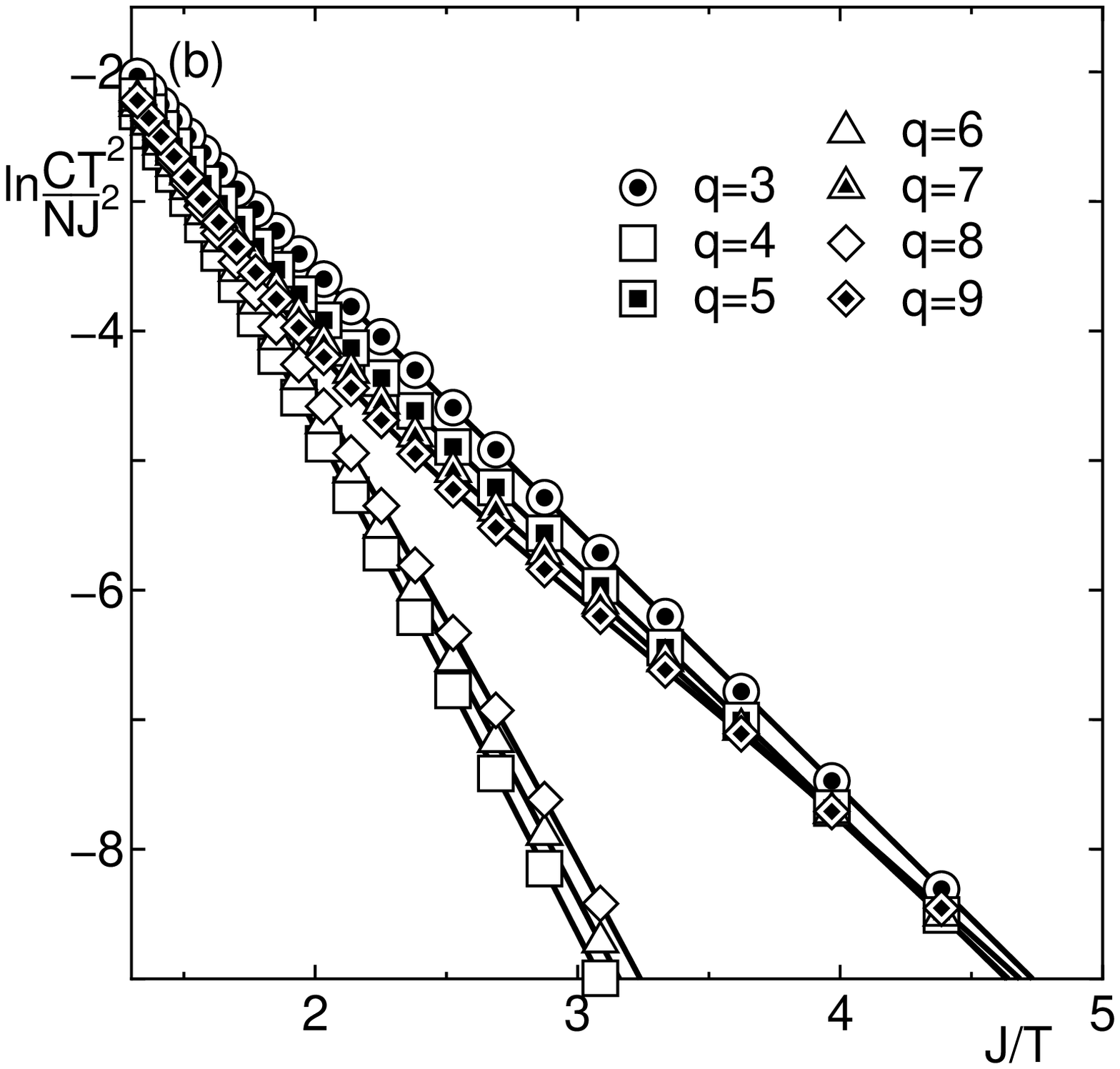}}
\caption{Arrhenius plot of the low temperature specific heat of (a) ladders and (b) tubes. In the inset of (a), the interval $2.5 \leq J/T \leq 3$ is magnified .}
\label{ahr}
\end{figure}
For the numerical calculation, we fix the weight of the boundary state as $x_{1,\alpha}=1/2^{q-1} (\alpha=1,...,2^{q-1})$ and generate $\v{x}_{n}$ and $\chi_{n}$ iterating (\ref{eq:x}) and (\ref{eq:chi}). After $n-1$ iterations we have $2^{(q-1)(n-1)}$ ($2^{q(n-1)}$) weight vectors $\v{x}_{n}$ on the $n$-th rung for ladders (tubes). For large enough $n$, they correspond to the bulk weight and the effect of the fixed boundary weight is washed out. Taking into account the self-averaging nature of $\ln\chi_n$, we obtain the thermodynamic limit of $F/N$ by averaging $-(T/q)\ln \chi_n$ over all possible $2^{(q-1)(n-1)}$  ($2^{q(n-1)}$) values of $\chi_n$.
 With the increase of  $\nrung$, however, the summation over all possible $\{t_{n,\alpha}\}$ becomes too demanding. Therefore, in our calculation, the average is taken over randomly chosen 1000 realizations of $\{t_{n,\alpha}\}$ with $200 \leq n \leq 1200$. 

  The specific heat and entropy are calculated by the numerical differentiation of the free energy. We calculate the free energy at different temperatures for the same set of $\{t_{n,\alpha}\}$. Since the free energy for each set of  $\{t_{n,\alpha}\}$ is a smooth function of temperature, the averaged free energy is also a smooth function of temperature. Therefore, the numerical differentiation can be  carried out without problem.

 The  specific heat $C$ and entropy ${\cal S}$ of ladders and tubes are plotted against $T$ in Figs. \ref{hine} and \ref{ent}, respectively.  For both ladders and tubes, the overall behavior is not sensitive to $q$ and is reminiscent of the two-dimensional $\pm J$ Ising model\cite{cm,kirk,mb,ba} except for the low temperature regime.
For tubes,  the entropy oscillates with the number of legs at low temperatures. This oscillation is not observed in ladders. This feature is also reflected in the low-temperature behavior of the specific heat as shown in the inset of Fig. \ref{hine}(b). 

To observe the $q$-dependence of the physical quantities in the low temperature limit clearly, the ground-state entropy is plotted against $q$ in Fig. \ref{entqdep}. The entropies of ladders and odd-leg tubes behave almost similarly, while the even-leg tubes have extra entropy in the ground state. The  ground-state energy also shows a similar  oscillation as shown in Figs. \ref{eneqdep}. These quantities are plotted against $1/q$ in Figs. \ref{entqdepinv} and \ref{eneqdepinv}.  The results for ladders and tubes approach the values for the two-dimensional $\pm J$ Ising model\cite{cm} plotted by the open right-directed triangles with the increase of $q$.

To observe the low-temperature asymptotic behavior of the specific heat, the quantity $\ln(CT^2/NJ^2)$ is plotted against $J/T$ in Fig. \ref{ahr}(a) for ladders and (b) for tubes.  The data for the ladders and the odd-leg tubes are close to each other and decrease with $q$. At low temperatures, they behave as $C T^2 \sim \exp(-2J/T)$ suggesting that the lowest excitation energy is $2J$. The data for the even-leg tubes are below those for ladders and odd-leg tubes. This is consistent with the result that the  ground-state entropies for the even-leg tubes are larger than those for other cases.  For the even-leg tubes, the specific heat increases with $q$. At low temperatures, they behave as $C T^2 \sim  \exp(-4J/T)$ suggesting that the lowest excitation energy is $4J$.

\begin{figure}
\centerline{\includegraphics[width=6cm]{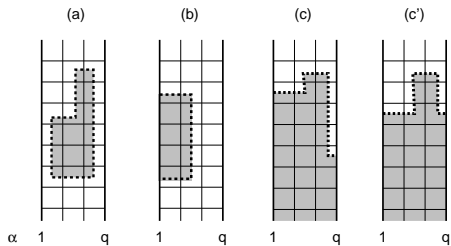}}
\caption{Types of excitations: (a) droplet, (b) edge droplet, (c) domain wall, and (c') closed domain wall compatible with the periodic boundary condition along the rungs. The spins in the shaded region are inverted relative to the ground-state configuration.}
\label{excite3}
\end{figure}
These features are understood by considering the elementary excitations of the present models, which can be classified into the following  three types:
\begin{enumerate}[{(}a{)}]
\item droplet excitation,
\item edge droplet excitation,
\item domain wall excitation.
\end{enumerate}
Schematic pictures of these excitations are given in Fig. \ref{excite3}. An excitation corresponds to the state with all spins in the shaded region inverted relative to the ground state. If an excitation has a vanishing excitation energy, it contributes to the ground-state degeneracy.

The excitation energy of a droplet excitation is a multiple of $4J$ including zero, because its boundary always contains even number of bonds. The edge droplet excitation is allowed only for the ladders and its  excitation energy is a multiple of $2J$. The  excitation energy of a domain wall is a multiple of $2J$ in the ladders. In the tubes, however, only the closed domain wall is compatible with the periodic boundary condition along the rungs as shown in Fig. \ref{excite3}(c'). In this case, the excitation energy of a domain wall is a multiple of $4J$ for the even-leg tubes and an odd-integer multiple of $2J$ for the odd-leg tubes. 
 Therefore, the lowest nonvanishing  excitation energy is $4J$ in even-leg tubes and $2J$ otherwise.

The above classification of excitations also helps to understand the excess ground-state entropy for the even-leg tubes. An even-leg tube can be formed by connecting two edges  ($\alpha=1$ and $q$) of an even-leg ladder. If the even-leg ladder has a domain wall with energy $2J$, we can connect its both ends  by inserting a vertical boundary between two edges.  Thus, the domain wall in the even-leg ladder is converted into a closed domain wall in the even-leg tube whose excitation energies are multiples of $4J$ including zero.   
The contribution from these zero energy excitations can be interpreted as the excess entropy.  
 It should be noted that this mechanism does not work for odd-leg cases, because the zero energy domain walls are not allowed in the odd-leg tubes. In the ladder, there is no constraint by the periodic boundary condition along the rungs. Hence, it is natural that the ground-state entropy and energy of the ladders vary smoothly with the number of legs.

\section{Summary and Discussion}

Finite temperature properties of the multileg $\pm J$ Ising ladders and tubes are investigated using the statistical transfer matrix method extending the method of Mattis and Paul\cite{mp}. It is found that the ground-state entropy shows an oscillating behavior with the number of legs in the tubes, while it decreases monotonically in the ladders. Corresponding behaviors of the specific heat and ground-state energy are found. From the numerical results for the specific heat, it is found that the lowest excitation energy  is $4J$ for the even-leg tubes, while it is $2J$ for other cases. The physical interpretation of these results is given by analyzing the structure of excited states.

The free energy of the two-leg ladder is calculated by MP using an approximate solution of their recursion relation. In the course of the present investigation, however, we found that it substantially deviate from our numerical solution for ladders with $q=2$. In addition, according to our numerical solution, the free energy does not approach the value of the ground-state energy predicted by MP. Actually, we found that the estimation of the ground-state energy by MP should be corrected. The corrected derivation of the ground-state energy and the numerical results for the temperature dependence of the free energy are given in Appendix.

Recent investigations for  the  two-dimensional $\pm J$ Ising model suggests the power law temperature dependence of the specific heat in spite of the finite energy gap.\cite{thm,jlmm}  This anomalous behavior is attributed to the presence of the infinite rigid spin cluster with fractal dimension. In the finite width ladders and tubes, the power law behavior is excluded at low temperatures as shown in Fig.\ref{ahr}. However, this figure also shows that the specific heat crosses over from  the high temperature regime, where the difference between the even-leg tubes and other cases is insignificant, to the low temperature regime, where this difference becomes significant. The crossover temperature decreases with the increase of $q$ suggesting the possibility that it tends to zero in the limit of $q\rightarrow \infty$. If this scenario is valid, the 'high temperature' regime can persist down to zero temperature in the limit $q\rightarrow \infty$ and the region with exponential temperature dependence would shrink to zero, allowing the power law behavior in the two-dimensional $\pm J$ Ising model.

There are many possible extensions of the present model. In general, the magnitudes of the rung and leg interactions should be taken unequal.  Similarly, the magnitudes of the ferromagnetic and antiferromagnetic interaction  should be unequal.  The probability of each type of bonds can be different. The quantum effect would be most important in application to the real ladder and tube materials at low temperatures. The investigation of these effects on the present model is left for future studies. 

The author thanks Y. Noguchi for collaboration in the early stage of this work. He also thanks  D. C. Mattis for suggestive comments to the earlier version of this work. This work is supported by a Grant-in-Aid for Scientific Research (C) (21540379) from Japan Society for the Promotion of Science.   The numerical computation in this work has been carried out using the facilities of the Supercomputer Center, Institute for Solid State Physics, University of Tokyo, Supercomputing Division, Information Technology Center, University of Tokyo, and   Yukawa Institute Computer Facility in Kyoto University. 
\appendix
\section{Ground State Energy of the $\pm J$ Ising Two-leg Ladder}

The ground-state energy  of the $\pm J$ Ising ladder has been calculated by Derrida {\it et al.}\cite{deri} for $q=3$ and Kadowaki {\it et al.}\cite{kado,ncom} for $q=2$ and $3$ using the zero temperature transfer matrix method. Here, we present a simple derivation for the case $q=2$ correcting the error of MP. 

When we assign  $\pm J$ 
 on the bonds of a ladder randomly, the probability that a plaquette consisting of four spins on two neighbouring rungs is frustrated (or unfrustrated) is  $1/2$. If we denote the frustrated and unfrustrated plaquette by F and U, respectively, each bond configuration is associated with a 
 series of letters F and U, which can be identified  
by numbers of successive  F and U as $\{n_{\rm U1},n_{\rm F1},n_{\rm U2},n_{\rm F2}...\}$. 
 We assume the first plaquette is U without affecting the conclusion in the thermodynamic limit. We call a cluster of plaquettes consisting of successive F's (U's)  bounded by U (F) on both sides, a F(U)-cluster.

In this representation, the ground-state energy for a bond configuration which  corresponds to the sequence $\{n_{{\rm U}i},n_{{\rm F}i}: i=1,N_c/2\}$ is given by
\begin{align}
E&=\sum_{i=1}^{N_c/2}\left\{E_{\rm U}(n_{{\rm U},i})+E_{\rm F}(n_{{\rm F,}i})\right\}
\end{align}
where $N_{\rm c}$ is the number of clusters.  $E_{\rm F}(n)$ and $E_{\rm U}(n)$ are the ground-state energies of F- and U-clusters with length $n$, respectively. For U-clusters, it is obvious that 
\begin{align}
 E_{\rm U}(n)=-(3n+1)J.
\end{align}
 For F-clusters,  it is energetically advantageous to put as many unsatisfied bonds as possible on the rungs which are shared by two F-clusters. This point was missed by MP.  
For even $n$, all unsatisfied bonds can be put on the rungs, while for odd $n$, one  unsatisfied bond must be on a leg.  Therefore,  we find 
\begin{align}
E_{\rm F}(n)=-\{2n-1-{\rm mod}(n,2)\}J.
\end{align}
Note that the energies of the rungs on the boundaries between F- and U-clusters are counted in $E_{\rm U}$. 

The total number of spins $N$ is given by 
\begin{align}
{N}
=2\sum_{i=1}^{N_c/2}\left(n_{{\rm U}i}+n_{{\rm F}i}\right).
\end{align}
Since $N$ is a macroscopic quantity, we can regard  $\aver{N}$ as the actual total number of spins. The probability that a sequence of $n$ letters appear is $1/2^n$. Hence, we find 
\begin{align}
\aver{n_{{\rm U}i}}=\aver{n_{{\rm F}i}}=\aver{n}=\sum_{n=1}^{\infty} \frac{n}{2^{n}}=2
\end{align} to obtain 
$\aver{N}=4N_c$. Similarly, the ground-state energy per spin is calculated as,
\begin{align}
\frac{\aver{E}}{N}
=\frac{N_cJ}{2N}\left\{-5\aver{n}+\aver{{\rm mod}(n,2)}\right\}=-\frac{7J}{6}\label{gene}
\end{align}
which is lower than the value $-J$ predicted by MP. 

\begin{figure}[h] 
\centerline{\includegraphics[width=6cm]{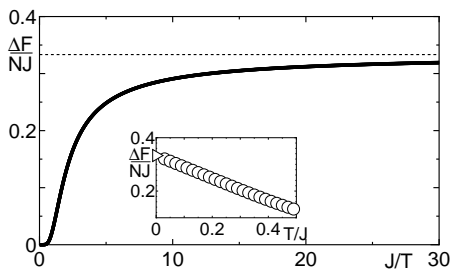}}
\caption{Excess free energy density $\Delta F/NJ$ (solid line) as a function
of $J/T$ calculated by direct numerical iteration of  (\ref{eq:x}) and (\ref{eq:chi}) for $\nrung=2$. The inset shows the low temperature behavior of $\Delta F/NJ$  plotted against $T/J$. The asymptotic approach to the exact ground-state value $1/3$ (dotted line; right-directed triangle of inset) is observed. }
\label{free}
\end{figure}

To check the consistency with the finite temperature calculation, we have estimated the free energy for $q=2$ by iterating (\ref{eq:x}) and (\ref{eq:chi}) for all possible realizations of  $\{t_{n,\alpha}\}$. 
  To make clear the comparison with Fig. 4 of MP, we plot the excess free energy $\Delta F \equiv F-F_{\rm pseudo}$ for $q=2$ against $J/T$ in Fig. \ref{free}, where $F_{\rm pseudo}$ is  the pseudoanealed free energy
\begin{align}
\frac{F_{\rm pseudo}}{N}=T\left\{\frac{1}{2}\ln 2 -\frac{3}{2}\ln\left(2\cosh \frac{J}{T}\right)\right\}
\end{align}
defined by MP.  The result substantially deviates from the approximate solution of MP. 
 Also, as $T$ tends to 0, it approaches $J/3$ as expected from  eq. (\ref{gene}) instead of $J/2$ which is predicted by MP.

\end{document}